\pdfoutput=1
\documentclass[a4paper,american,aip,citeautoscript,floatfix,pdftex,jcp,showpacs,superscriptaddress,twoside,%
longbibliography,%
preprint%
]{revtex4-1}
\usepackage{amsfonts,amsmath,amssymb}
\usepackage{graphicx}
\usepackage[T1]{fontenc}
\usepackage[utf8]{inputenc}
\usepackage{microtype}
\usepackage{textcomp}
\usepackage[textsize=tiny,obeyFinal]{todonotes}
\usepackage{xspace}
\usepackage{hyperref}
\usepackage{hypernat}
\usepackage[todos]{CMImacros}

\newlength{\figwidth}
\newcommand{\cfeldesy}{\affiliation{Center for Free-Electron Laser Science, Deutsches
      Elektronen-Synchrotron DESY, Notkestrasse 85, 22607 Hamburg, Germany}}%
\newcommand{\uhhcui}{\affiliation{The Hamburg Center for Ultrafast Imaging, Universit\"at Hamburg,
      Luruper Chaussee 149, 22761 Hamburg, Germany}}%
\newcommand{\uhhchem}{\affiliation{Department of Chemistry, Universit\"at Hamburg,
      Martin-Luther-King-Platz 6, 20146 Hamburg, Germany}}%
\newcommand{\uhhphys}{\affiliation{Department of Physics, Universit\"at Hamburg, Luruper Chaussee
      149, 22761 Hamburg, Germany}}%

\begin{document}
\title{Characterizing and optimizing a laser-desorption \mbox{molecular beam source}}%
\author{Nicole\ Teschmit}\cfeldesy\uhhcui\uhhchem%
\author{Karol D\l{}ugo\l{}\c{e}cki}\cfeldesy%
\author{Daniel Gusa}\cfeldesy%
\author{Igor Rubinsky}\cfeldesy%
\author{Daniel A. Horke}\cfeldesy\uhhcui%
\author{Jochen Küpper}\email{jochen.kuepper@cfel.de}%
\homepage{https://www.controlled-molecule-imaging.org}%
\cfeldesy\uhhcui\uhhchem\uhhphys%
\date{\today}%
\begin{abstract}\noindent%
   The design and characterization of a new laser-desorption molecular beam source, tailored for use
   in x-ray-free-electron-laser and ultrashort-pulse-laser imaging experiments, is presented. It
   consists of a single mechanical unit containing all source components, including the
   molecular-beam valve, the sample, and the fiber-coupled desorption laser, which is movable in
   five axes, as required for experiments at central facilities. Utilizing strong-field ionization,
   we characterize the produced molecular beam and evaluate the influence of desorption laser pulse
   energy, relative timing of valve opening and desorption laser, sample bar height, and which part
   of the molecular packet is probed on the sample properties. Strong-field ionization acts as a
   universal probe and allows to detect all species present in the molecular beam, and hence enables
   us to analyze the purity of the produced molecular beam, including molecular fragments. We
   present optimized experimental parameters for the production of the purest molecular beam,
   containing the highest yield of intact parent ions, which we find to be very sensitive to the
   placement of the desorbed-molecules plume within the supersonic expansion.
\end{abstract}
\maketitle

\section{Introduction}
Laser desorption (LD) is a widely used technique to vaporize non-volatile organic molecules for
gas-phase studies. The concept of LD is a rapid heating of the sample to be vaporized, at around
$10^{10}\text{--}10^{12}~\text{K/s}$, such that a fraction of molecules desorb intact instead of
fragmenting~\cite{Vastola:AMS4:107, Rijs:Springer:1}. Later studies combined LD with pulsed
molecular beams to directly cool the desorbed molecules, enabling the investigation of intact
neutral molecules in the gas-phase at low vibrational temperatures~\cite{Tembreull:AnalChem59:1003,
   Meijer:APB51:395, Nir:Nature408:949, DeVries:ARPC58:585}. The main advantage of LD over other
vaporization techniques, such as thermal vaporization, is the ability to introduce intact thermally
labile organic molecules, including peptides and proteins, into a cold molecular beam. This has been
demonstrated, \eg, for a pentapeptide (Ser-Ile-Val-Ser-Phe-NH$_2$)~\cite{Ishiuchi:PCCP18:23277} or
the delta sleep inducing nonapeptide~\cite{Bakker:CPC:CPHC6:120}.

A first detailed characterization of a LD source coupled to a molecular beam (MB) valve was
conducted nearly 30 years ago. Using anthracene, diphenylamine, and perylene, combined with
resonance-enhanced multiphoton ionization (REMPI) spectroscopy, approximate vibrational temperatures
of $<\!15$~K and rotational temperatures of 5--10~K were determined~\cite{Meijer:APB51:395}. This
demonstrated the ability of LDMB to gently vaporize large, thermally labile molecules and to
efficiently cool them. Since then various spectroscopic techniques have been combined with LDMB
sources and recent experiments have included resonance-enhanced multiphoton ionization
studies~\cite{Nir:Nature408:949}, (far) infrared (IR)-ultraviolet (UV) double resonance
techniques~\cite{Bakker:PRL91:203003}, IR multiphoton dissociation~\cite{Ligare:JPCB119:7894}, and
zero-kinetic-energy-photoelectron (ZEKE) spectroscopy~\cite{Zhang:JCP128:104301}.

In recent years, x-ray free-electron lasers (XFELs) have emerged as powerful tools for
structure determination of gas-phase systems, with the potential to achieve atomic-resolution
structures with femtosecond temporal resolution, recording so-called \emph{molecular
movies}~\cite{Barty:ARPC64:415}. The ultrashort pulse duration available at XFELs enables the
recording of a diffraction pattern from a molecule prior to destruction by the high intensity of the
x-ray pulses~\cite{Neutze:Nature406:752}. This \emph{diffraction-before-destruction} paradigm,
albeit still discussed~\cite{Ziaja:NJP14:115015, Lorenz:PRE86:051911, Nass:JSR22:225}, has recently
also been demonstrated for isolated gas-phase molecules~\cite{Kuepper:PRL112:083002,
Stern:FD171:393, Glownia:PRL117:153003}. Similar to the time-resolved nuclear dynamics that can be
recorded at XFEL sources, modern laboratory based attosecond light sources allow the measurement of
real-time electron dynamics in isolated molecules~\cite{Calegari:Science346:336}.

These experiments, however, are themselves inherently not species specific, \ie, all molecules
within the interaction region will be probed. Therefore, combination of LD with XFEL and attosecond
experiments requires a pure molecular sample in the gas-phase. Furthermore, to be compatible with
central facility light-sources, the laser desorption source needs to be translatable in three axes
to adjust the molecular beam to the fixed XFEL beam. Additionally, the continuous measurement time
should be as long as possible and the sample quickly exchangeable.

Here, we detail the characterization, and optimization of our novel LDMB source
   design, constructed to be compatible with central facilities, such as XFELs or attoscience
centers. Using the dipeptide Ac-Phe-Cys-NH$_2$ as a prototypical labile biological molecule, which
has first been laser desorbed and studied by the Rijs
group~\cite{Yan:PCCP16:10770, Alauddin:PCCP17:2169}, we characterize the created beam using
strong-field ionization with a femtosecond laser pulse. This allows us to monitor all
   species present in the interaction region, including the carrier gas of the supersonic
   expansion. We show the optimization of experimental parameters to reduce fragmentation, to
improve cooling of desorbed molecules and, thereby, to maximize the phase-space density of intact
parent molecules in the interaction region. The created molecular beams are well-suited to further
manipulation and purification, \eg, using electrostatic deflection
techniques~\cite{Chang:IRPC34:557}, an important step towards recording temporally and spatially
resolved nuclear and electronic dynamics of isolated biomolecules.

\section{Experimental Setup}
\label{sec:setup}
\begin{figure}
  \centering
  \includegraphics[width=\figwidth]{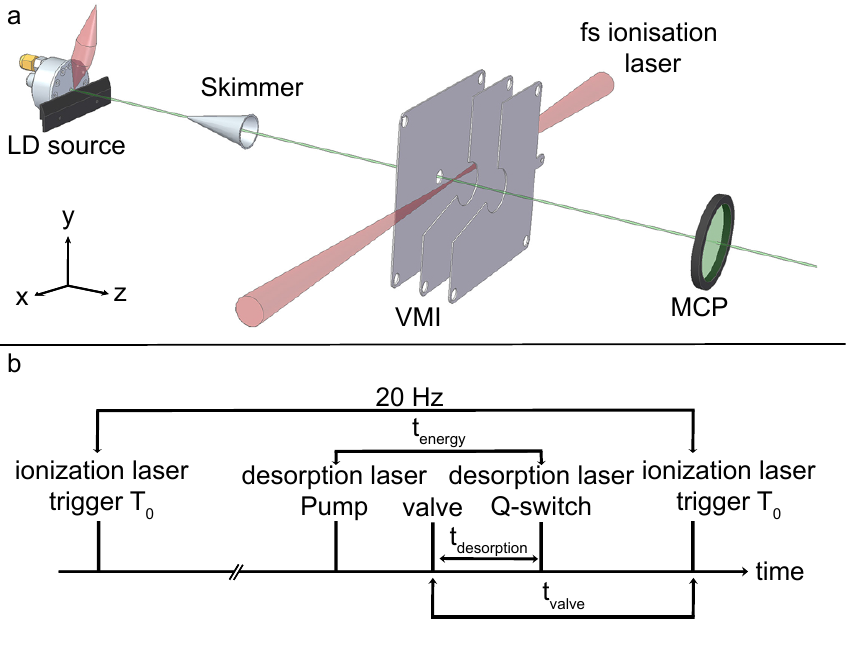}
  \caption{(a) Experimental setup for characterizing a laser desorption source. The source chamber
     contains the laser desorption source and and is separated from the detection chamber by a 2~mm
     conical skimmer. The detection chamber houses a velocity-map imaging setup and microchannel
     plate detector. (b) Experimental timing diagram. The master trigger is given by the ionization
     laser, a kHz-repetition-rate femtosecond-pulse laser system that cannot trivially
        be externally triggered, and the molecular beam valve is triggered relative to this with
     delay $t_\text{valve}$. The desorption laser trigger is defined relative to the valve with
     delay $t_\text{desorption}$, while the energy of the desorption laser can be changed by
     modifying the Q-switch timing $t_\text{energy}$}
  \label{fig:setup}
\end{figure}

The mechanical design and construction of this LD source is based on compatibility with
large-scale-facility-based photon sources. The laser-desorption source consists of a single central
mechanical unit containing all necessary parts (molecular beam valve, sample bar with motors, and
desorption laser optics). It is mounted on a three axis manipulator on a single flange for
independent motion in the source chamber, which is pumped with a turbo molecular pump (Pfeiffer
Vacuum HiPace 700P) to operating pressures typically around $10^{-5}$~mbar. It contains a cantilever
piezo valve~\cite{Irimia:RSI80:113303} operated at 6~bar backing pressure of argon. The valve has a
300~\um orifice, followed by a conical opening of 4~mm length and 40$^{\circ}$ opening angle.
 Conical nozzle shapes are well known to produce molecular beams with more efficient
   translational cooling and greater directionality, and hence densities, than simple pinhole
   sources~\cite{Luria:JPCA115:7362}. A graphite (Poco EDM-1) sample bar (80~mm long, sample
channel width 1.2~mm) is placed approximately 200~\um in front of the valve, see
\autoref{fig:setup}. The sample bar height ($y$~axis) can be translated using an in-vacuum two-phase
stepper motor (Owis SM.255.V6). To replenish the molecular sample, the sample bar can be moved along
the $x$-direction using an in-vacuum linear piezo-stage (SmarAct SLC-24120-S-HV), typically operated
at 0.02~mm/s. This results in measurement times of around 70~min per sample bar. For longer
measurement periods the sample bar can be quickly exchanged with a load-lock system, pumped by a
Pfeiffer Vacuum HiCube 80 Eco pumpstand (typical turn-around time 10~minutes). The entire molecule
source (valve, sample-bar holder with motors, and desorption-laser optics) is placed on a three axis
manipulator and can furthermore be adjusted for tip and tilt angle, allowing independent five-axes
motion of the device within the vacuum chamber, as required for experiments at XFEL facilities. It
is generally useful for operation of the source in molecular-beam setups were accurate alignment of
the source is crucial, \eg, multi-skimmer setups or electrostatic manipulation
devices~\cite{Chang:IRPC34:557, Bethlem:JPB39:R263}. Detailed drawings of the source
   and individual components are given in the supplementary information.

Molecules on the sample bar (see below for sample preparation procedures) are desorbed by pulses
from a fiber-coupled, diode-pumped Nd:YAG laser at 1064~nm (Innolas Spitlight Compact DPSS10),
operating at 20~Hz with a pulse duration (full width at half maximum) of 9~ns and pulse energy up to
0.8~mJ. This is coupled into a multimode fiber (CeramOptec WF 400/440P) with core diameter 400~\um
and numerical aperture of 0.22. The fiber is coupled into the vacuum chamber with a custom-made
Swagelock connection and polytetrafluoroethylene (PTFE) plug~\cite{Abraham:AO37:1762}. Inside the
chamber the fiber is out-coupled with a custom-made vacuum compatible fiber collimator and the laser
beam is focused to a spot size of approximately 0.6~mm on the sample bar. Custom mounting of the
collimator allows variation of the laser spot size on the sample, as well as translation of the
laser beam along the $x$ and $z$ axes, and tilting in the $yz$ plane.

Following desorption, molecules are picked up by the supersonic argon jet from the valve and rapidly
cooled down. The resulting molecular beam is skimmed with a 2~mm diameter skimmer (Beam Dynamics
Inc.\ Model 50.8), located approximately 5~cm downstream the valve. Following the skimmer the
molecular beam enters the differentially pumped (Pfeiffer Vacuum HiPace 2300) detection chamber,
maintained at pressures around $3\times10^{-7}$~mbar. The detection chamber contains a velocity-map
imaging (VMI) setup with a classic Eppink and Parker 3-plate design~\cite{Eppink:RSI68:3477}. The
distance from the molecular beam valve to the interaction point is around 45~cm. For the results
presented here, the VMI setup was operated as a time-of-flight mass spectrometer, with typical mass
resolution $m/\Delta{m}\approx100$.

\begin{figure*}
   \centering
   \includegraphics[width=170mm]{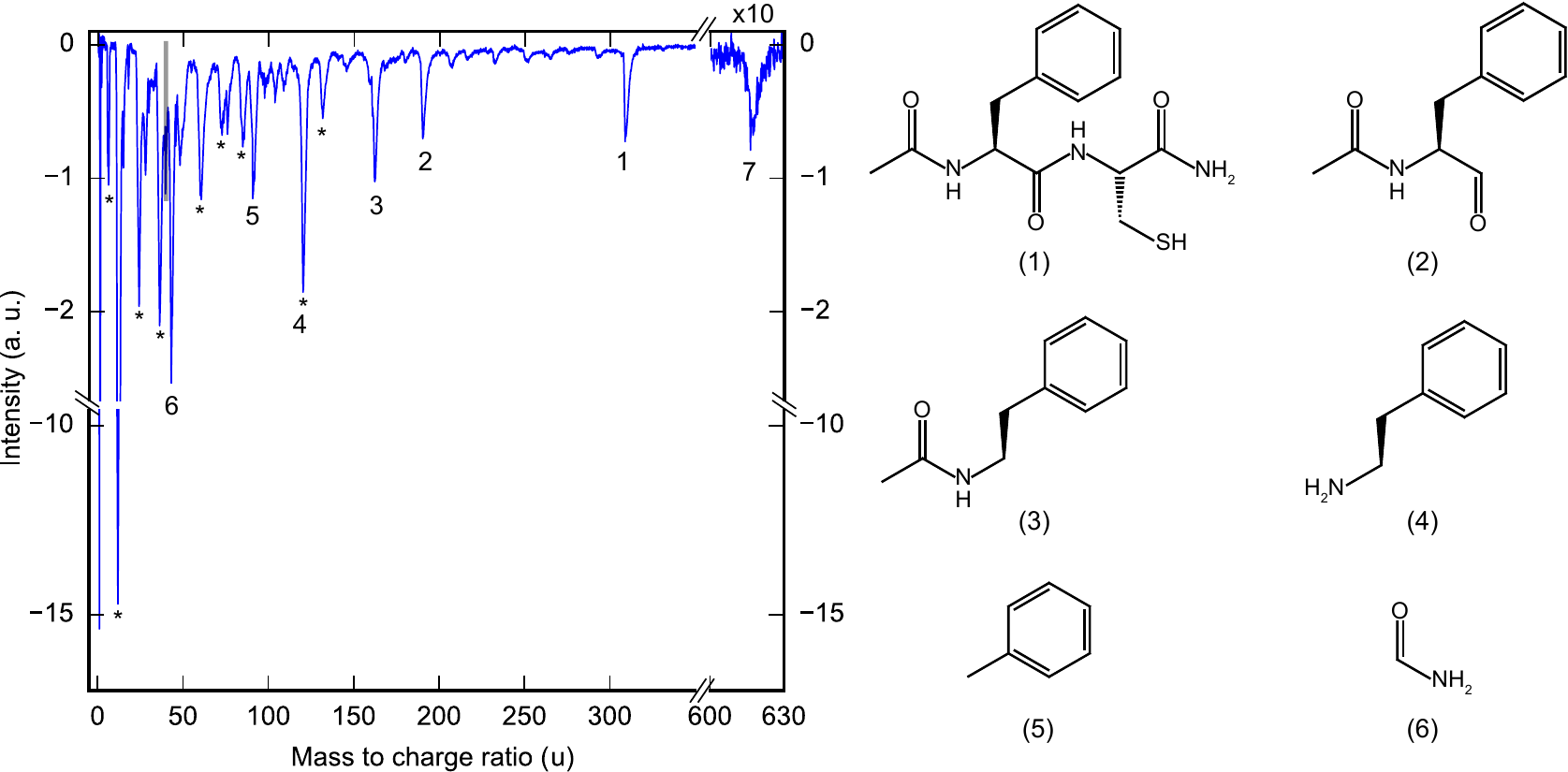}
   \caption{Time-of-flight mass spectrum of laser desorbed APCN following strong-field ionization.
      Peaks resulting from carbon or carbon clusters are labeled with $\ast$. 1 corresponds to the
      parent ion, and 2--6 to molecular fragments. A small signal from APCN dimer is also observed,
      7. Grey shading marks the peak from argon carrier gas, which appears much smaller than
      molecular fragments due to the much higher ionization potential.}
   \label{fig:TOF}
\end{figure*}

The molecular beam is probed \emph{via} strong-field ionization using a femtosecond Ti:Sapphire
laser system (Spectra Physics Spitfire Ace) with a central wavelength of 800~nm, a pulse duration of
40~fs and pulse energies up to 300~\uJ. It is focused into the vacuum chamber with a $f=800$~mm lens
to a spotsize (FWHM) of 80~\um in the interaction region between the VMI electrodes.

The timing scheme for our experimental setup is shown in \autoref{fig:setup}. The femtosecond
ionization laser is used as the master trigger in the experiment. Its native repetition rate of
1~kHz is electronically reduced to provide a trigger signal at 20~Hz. The molecular beam valve
trigger is defined, relative to this fs-laser trigger, by the delay $t_\text{valve}$. This delay
defines which (longitudinal) part of the molecular beam we are probing with the ionization laser.
The desorption laser is now triggered relative to the valve trigger and defined by the delay
$t_\text{desorption}$, which controls where within the gas pulse the desorbed molecules are placed.
This setup enables us to change $t_\text{valve}$ without changing $t_\text{desorption}$. We note
that the timing values given should be seen as relative, not absolute values, as they are
susceptible to electronic delays within the valve and laser controls used.

The velocity of the molecular beam is measured by recording the temporal profile of the beam, \ie,
scanning $t_\text{valve}$, for different longitudinal positions of the valve, \ie, its distance
from the first skimmer. We then evaluate the beam velocity from the temporal shift in the peak of
the parent ion for different valve positions to be approximately 670~m/s.

The dipeptide Ac-Phe-Cys-NH$_2$ (APCN, 95\% purity, antibodies-online GmbH) is used in this study
without further purification. The sample powder is mixed with graphite powder (0.44:1 by weight) and
ground with pestle and mortar to a fine powder. The top surface of the graphite sample bar is
roughened with sand paper and pushed into the prepared sample mixture. Gentle force is used to
ensure the mixture sticks to the sample bar and an even sample layer is formed.

\section{Results and Discussion}
\label{sec:results}
\subsection{Molecular fragmentation}
A measured time-of-flight mass spectrum of laser desorbed and strong-field ionized APCN is shown in
\autoref{fig:TOF}. It shows clear signals from parent ions (1 in \autoref{fig:TOF}), fragments ions
(2--6) and parent dimer (7). Furthermore, we observe several peaks from carbon and carbon clusters
consisting of up to 11 atoms (highlighted with an asterisk in \autoref{fig:TOF}). These are present
due to their direct desorption from the graphite matrix material within which the sample is
embedded, as well as due to the formation of higher-order carbon clusters within the desorption
plasma created by the laser pulse~\cite{Rohlfing:JCP81:3322}. The molecular fragments originating
from the APCN sample identified in the spectrum are shown in \autoref{fig:TOF}. Strong-field
ionization is a non species-selective method and thus allows the identification of all species
present within the molecular beam. This approach, therefore, allows us to optimize the yield and
fraction of intact parent molecules contained within the molecular beam. For further analysis of the
contributing parameters for laser desorption, we identify four characteristic fragments; the APCN
parent ion ($m/z=309$, peak labeled 1), the C$_{10}$H$_{12}$NO fragment ion ($m/z=162$, 3), the
argon ion peak ($m/z=40$), and the carbon peak ($m/z=12$). The particular molecular fragment (3) is
chosen as it provides the largest-intensity clean signal, \ie, it does not overlap with a carbon
cluster fragment.

\begin{figure*}
   \centering
   \includegraphics[width=6in]{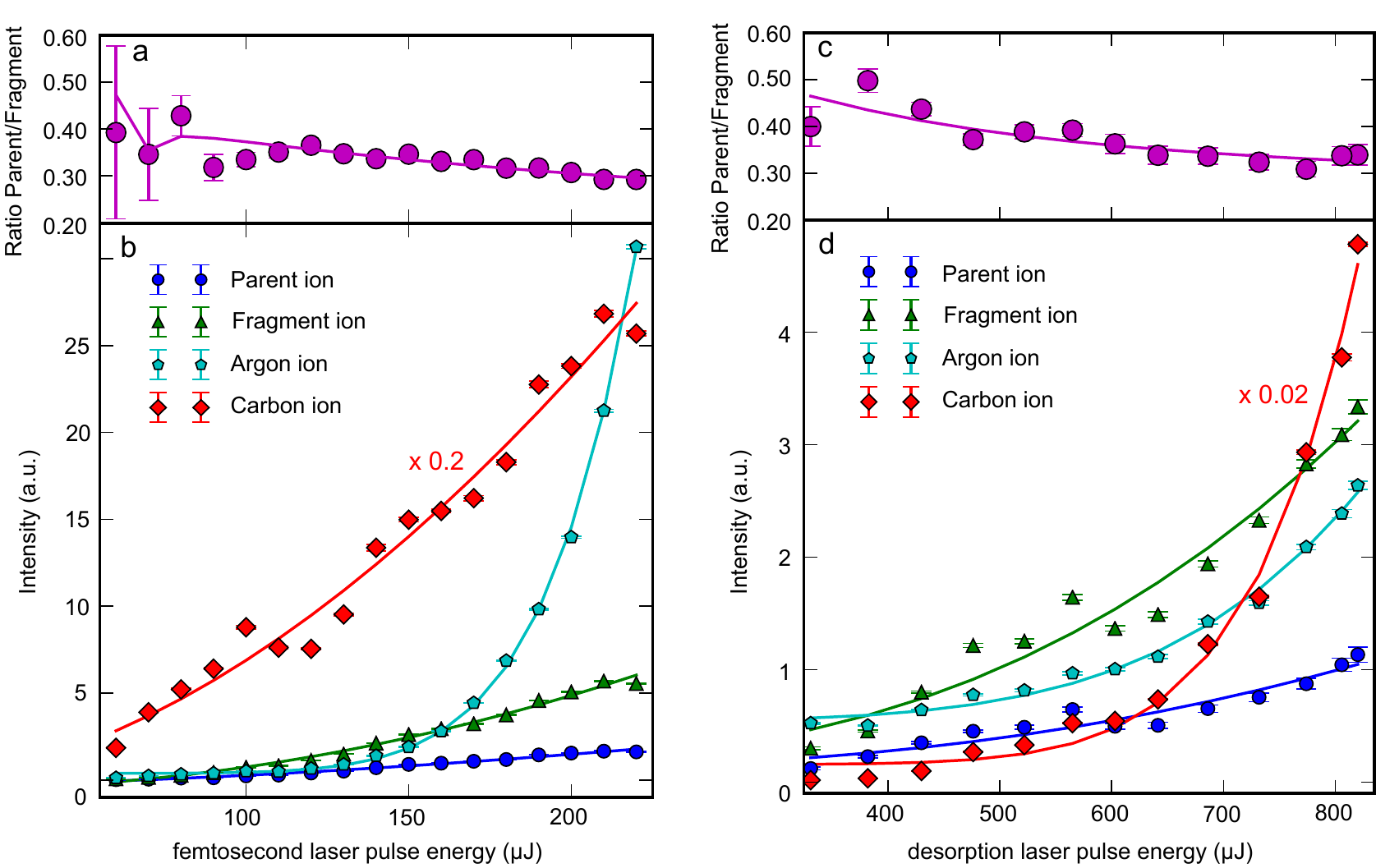}
   \caption{Measured ion intensity for parent APCN, the characteristic fragment at $m/z=162$, carbon
      and argon as a function of ionization laser pulse energy (a,b) and desorption laser pulse
      energy (c,d). Solid lines are power law fits to the data. The top panels (a,c) show the ratio
      of parent to fragment ions observed. Throughout the manuscripts, shown error bars correspond
      to 1~standard error (std.~err.).}
   \label{fig:laser-power}
\end{figure*}

In order to evaluate the effect of the femtosecond ionization laser on the observed molecular
fragmentation, we scan the laser pulse energy between 60~\uJ and 220~\uJ. The observed integrated
ion intensities for the 4 characteristic peaks are shown in \autoref{fig:laser-power}~b. The solid
lines are a power-law fit of the form $A\times x^{n}+c$. Additionally, in panel (a) we plot the
ratio of observed parent ions to the selected fragment ions for the probed intensity region. This
nearly constant ratio indicates that the strong-field ionization process has little influence on the
fragmentation of the parent ion. This is in agreement with previous studies that indicated that the
fragmentation of complex molecules in intense laser fields is very sensitive to the laser pulse
duration, but not the total energy of the pulse~\cite{Calvert:PCCP14:6289}. Therefore,
increasing the ionization laser intensities leads to larger ion signals, but has little effect on
the fragmentation patterns observed. Hence using strong-field ionization with ultrashort laser
pulses is a powerful tool for the full characterization of molecular beams containing complex
molecules and fragments thereof.

To elucidate the effect of the high-power desorption laser on the molecular sample on the graphite
sample bar, we record mass spectra for various desorption laser energies in the range
$\sim350$--800~\uJ. In \autoref{fig:laser-power}~d, we plot the recorded integrated ion intensities
for the four characteristic masses as a function of desorption laser energy, with solid lines
corresponding to a power law fit to the data. As can be seen from these data, all intensities
increase with increasing laser energy. This includes, somewhat un-intuitively, the argon seed gas
signal observed in the interaction region, which will be discussed later on. The carbon signal shows
the steepest dependence on desorption laser pulse energy, which is consistent with the formation of
isolated carbon atoms and clusters within a laser-induced plasma~\cite{Rohlfing:JCP81:3322}.
\autoref{fig:laser-power}~c shows the ratio of parent to fragment signal. A decrease in the
parent-to-fragment ratio is observed as the desorption energy is increased. Thus the desorption
process can cause fragmentation of the sample and the highest fraction of parent ion within the
molecular beam is obtained at the lowest desorption energies, albeit at the expense of density
within the beam. We furthermore note that the actual fractional yield of parent ions within the
molecular beam is significantly smaller than the numerical values shown in
\autoref{fig:laser-power}, since these only take into account a single characteristic fragment.
Additionally, any charged fragments produced during the desorption process will not arrive in the
detection region due to the static fields applied to the time-of-flight electrodes. All studies
below were conducted with around 670~\uJ desorption laser pulse energy and 140~\uJ ionization laser
power.

\subsection{Molecular beam properties}
\label{sec:mol_beam}
\begin{figure}
   \centering
   \includegraphics[width=\figwidth]{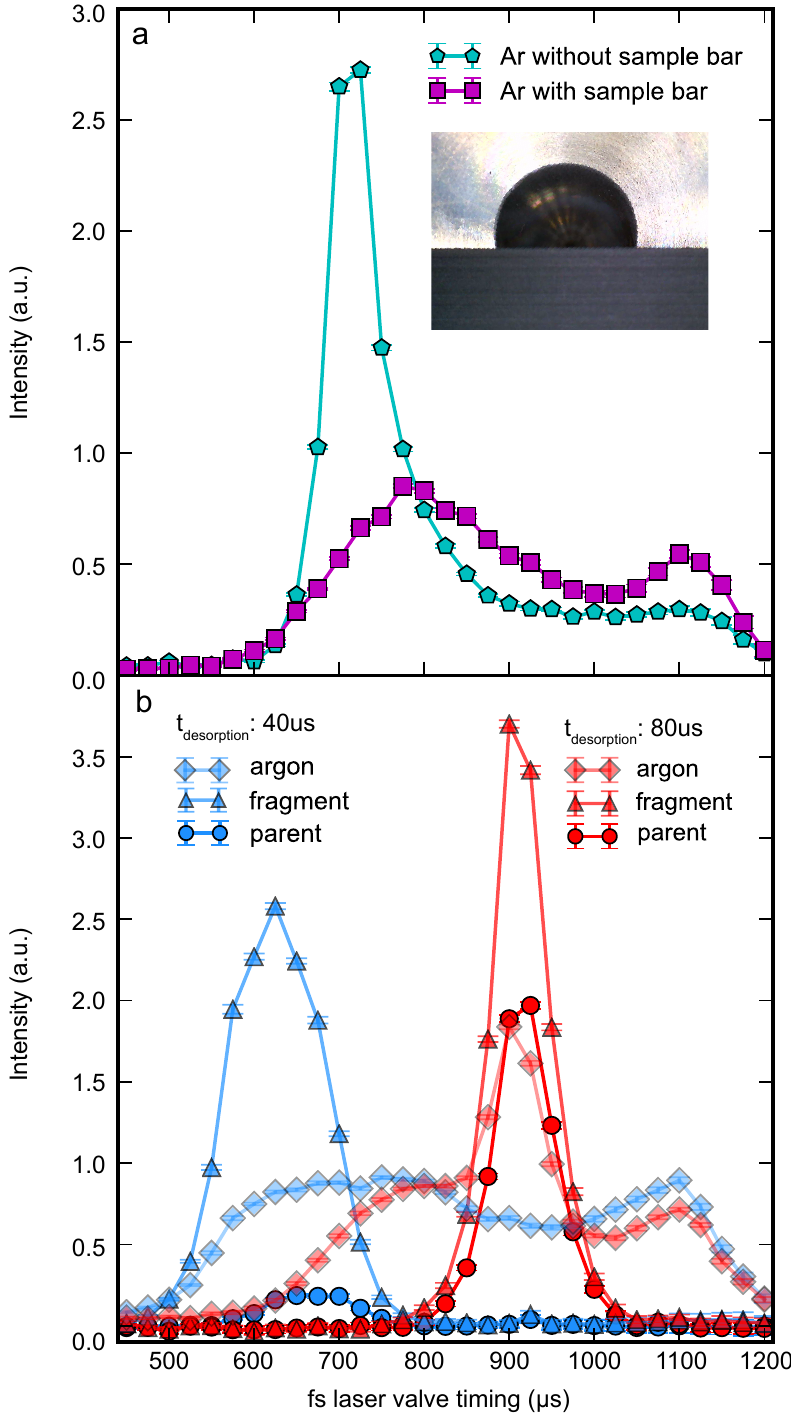}
   \caption{Integrated ion intensity as a function of valve delay $t_\text{valve}$, \ie, temporal
      (longitudinal) profiles of the molecular beam. (a) argon profile with (purple squares) and
      without (torquois pentagons) sample bar in front of the valve. The inset shows a photograph of
      the sample bar in front of the valve at the optimized position; see text for details. (b)
      Temporal profile for APCN parent, characteristic fragment and argon for two different
      desorption laser delays, $t_\text{desorption}$.}
   \label{fig:beam_timing}
\end{figure}
To probe the longitudinal (or temporal) profile of the produced molecular beam, we scan the timing
between the valve trigger and the ionization laser, $t_\text{valve}$, see \autoref{fig:setup},
probing different portions of the molecular beam. A typical temporal profile is shown in
\autoref{fig:beam_timing}~a for the pure argon beam emerging from the valve with (purple) and
without (blue) the graphite sample bar in place. Without the sample bar, we observe a single sharp
peak with a full width at half maximum of $\sim75$~\us (corresponding to a speed ratio of $\sim$10),
and a small shoulder at longer times due to the rebounce of the piezo within the
valve~\cite{Irimia:RSI80:113303}. When the sample bar is placed in front of the valve, as shown in
the picture inset in \autoref{fig:beam_timing}, the argon gas flow is significantly disturbed. The
overall gas-pulse is significantly broader, the main peak intensity is decreased by a factor of
$\sim$3, and more intensity is observed at later times. We attribute these observations to the
disturbance of the argon flow by the sample bar and possible turbulences in the flow-field within
the dead volume behind the sample bar.

By changing the relative timing of the desorption laser and valve trigger ($t_\text{desorption}$ in
\autoref{fig:setup}), we can now place the plume of desorbed molecules at different positions within
the argon beam. Changing this relative timing has significant effects on the observed intensities of
argon, parent, and fragment ions alike, as shown in \autoref{fig:beam_timing}~b. We show temporal
molecular beam profiles for argon (diamond markers), fragment 3 (triangles), and parent ions
(circles) at two different time delays $t_\text{desorption}$. The relative timing of the desorption
laser and molecular beam valve significantly affects the intensity of fragment and parent ions, and
the ratio between the two. We attribute this effect to changes in the cooling efficiency as the hot
desorbed molecules are placed within the argon expansion at different times. An efficient cooling
process is required to quench the excess energy of the desorbed molecules and prevent further
fragmentation. Comparing the relative integrated intensities of parent and fragment signals at the
two timings shown in \autoref{fig:beam_timing}~b, we observe that the combined intensity is
approximately identical at the two time points, however the ratio between the two differs
significantly. This suggests that while approximately the same density of molecules, including
fragments, is present within the initially desorbed plume, the less efficient cooling in the
less-dense front of the gas pulse at the delay of 40~\us leads to significant fragmentation
occurring before or during the argon gas pulse, \ie, the cooling occurs too late and fragmentation
has already taken place.

Further to its influence on the molecular signals the timing of the desorption laser clearly has an
effect on the observed argon signal. An increase in argon signal is observed at the timing where
desorbed molecules (parent or fragments) are present within the beam. We suspect the observed
increase in the argon signal is not due to more argon arriving at the detector, but due to signals
of molecular fragments or carbon clusters overlapping with the argon mass at 40~u in a very crowded
spectrum shown in \autoref{fig:TOF}.

\begin{figure}
   \centering
   \includegraphics[width=\figwidth]{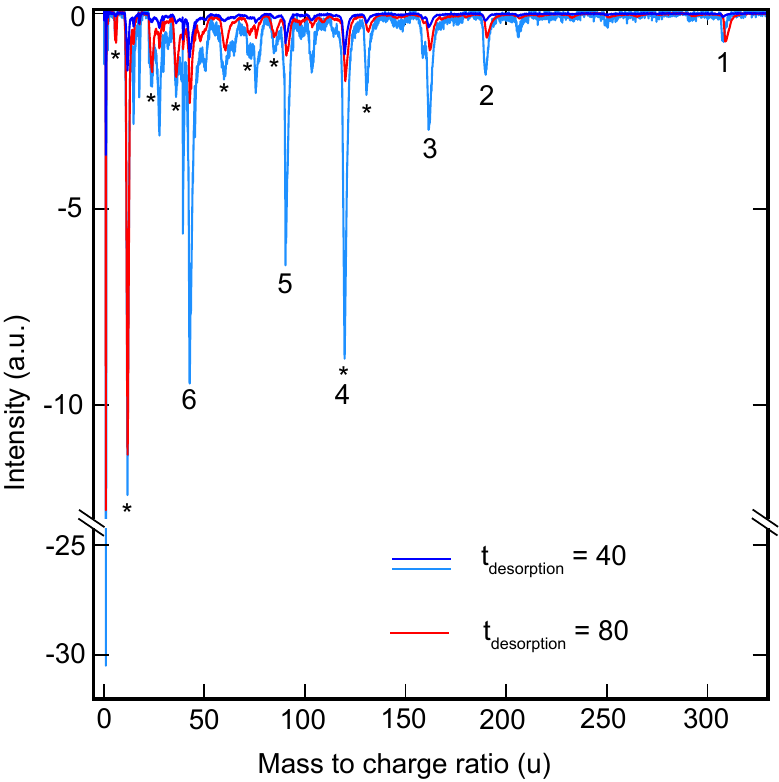}
   \caption{Mass spectrum for the two different desorption laser timings shown in
      \autoref{fig:beam_timing}(b), red and dark blue are the absolute intensities measured, the
      light blue trace has been scaled to the parent ion intensity of the red
      ($t_\text{desorption}=80$~\us) trace. A significant increase in fragment ion yield is observed
      for $t_\text{desorption}=40$~\us.}
   \label{fig:tof_comparison}
\end{figure}
The dependence of the observed fragmentation on the relative position in the gas pulse is,
furthermore, evident when comparing mass spectra recorded at different $t_\text{desorption}$, as
shown in \autoref{fig:tof_comparison}. These spectra have been recorded with $t_\text{valve}$
optimized for parent ion signal and are plotted normalized with respect to the observed parent
intensity. This shows that all molecular fragments are significantly more abundant at
$t_\text{desorption}=40~\us$, indicating a much higher internal temperature of the desorbed
molecules, due to the less efficient cooling, and correspondingly increased fragmentation.

\begin{figure}
   \centering
   \includegraphics[width=85mm]{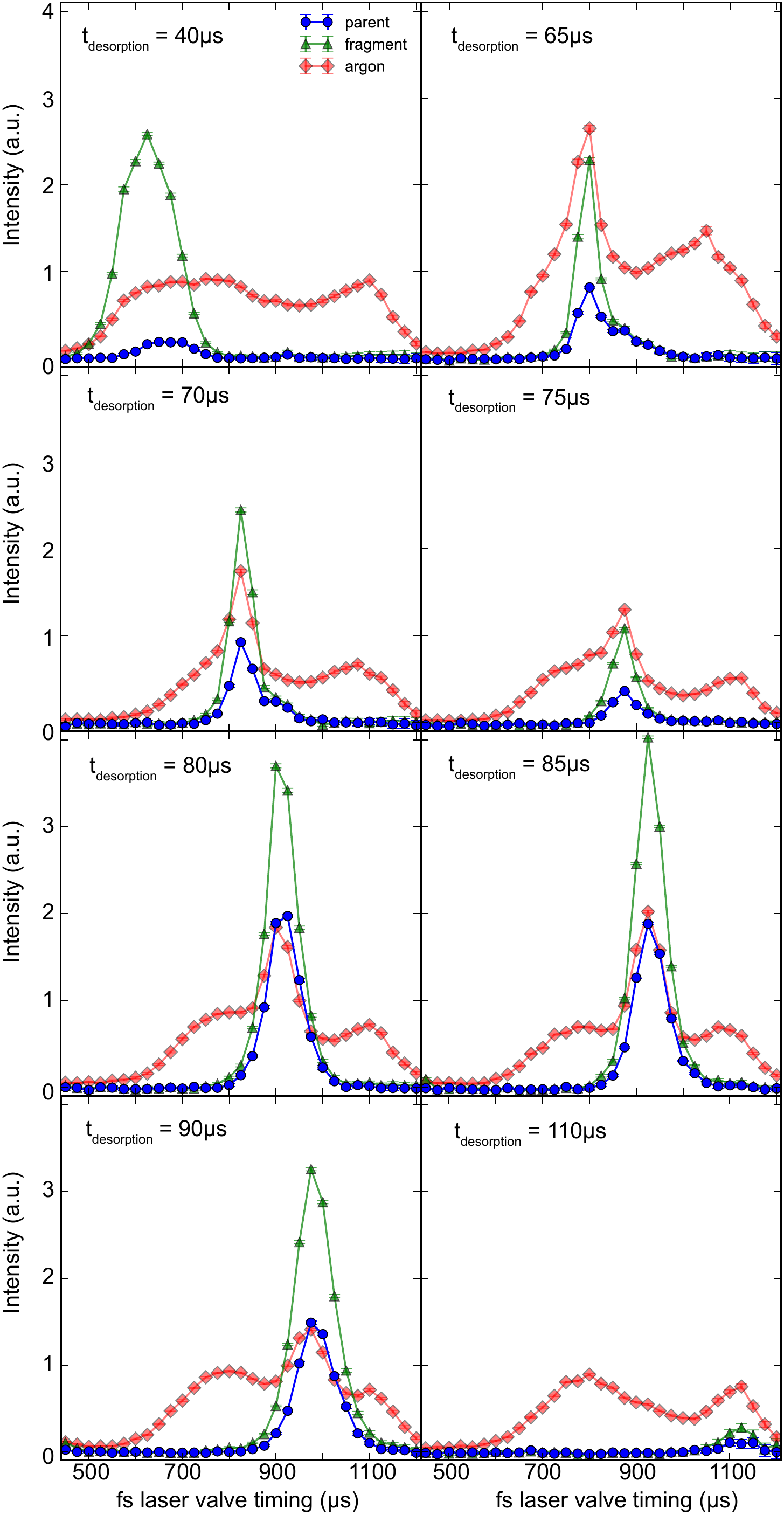}
   \caption{Longitudinal (temporal) profiles of the molecular beam for different
      $t_\text{desorption}$ delays. Shown are intensities for the intact parent (blue circles),
      fragment (green triangles) and argon (red squares) ions.}
   \label{fig:waterfallplot}
\end{figure}

\begin{figure}
   \centering
   \includegraphics[width=\figwidth]{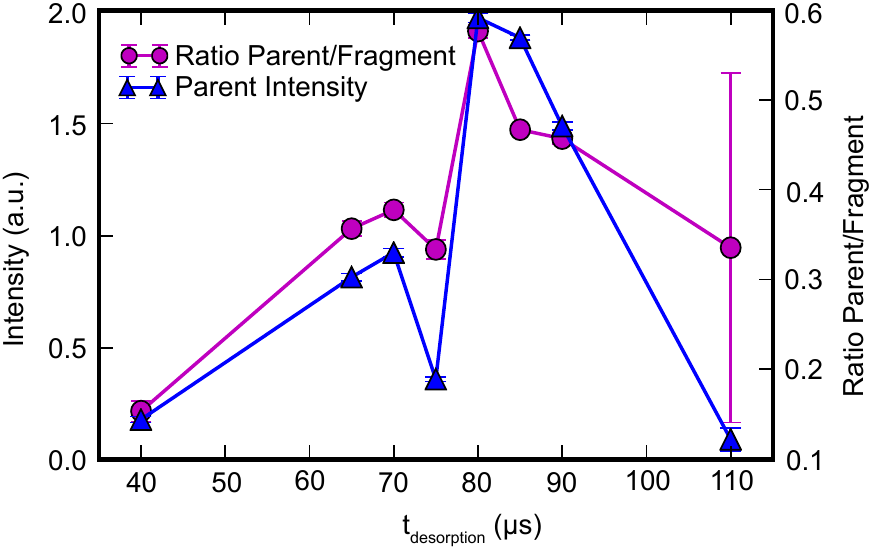}
   \caption{Dependence of the parent ion intensity and the parent to fragment ratio on the
      desorption laser timing $t_\text{desorption}$; error bars correspond to 1 standard error.
      Optimum conditions are observed for $t_\text{desorption}=80$~\us. The data at 75~\us shows a
      significantly lower intensity than expected, which could be due to irregularities in sample
      bar preparation; see text for details.}
   \label{fig:peak_ratio}
\end{figure}

To identify the optimum placement of the desorbed plume within the gas pulse, we have repeated these
measurements for several delay points, shown in \autoref{fig:waterfallplot}, where the curve with
higher intensity always corresponds to the fragment and the curve with lower intensity to the parent
ion.
For better comparison the parent intensity for each $t_{desorption}$, as well as the ratio of parent
to the characteristic fragment, is shown in \autoref{fig:peak_ratio}.

For our setup and under the given experimental conditions, we identify a delay of 80~\us as
providing the highest total intensity of parent signal, as well as the best parent-fragment ratio.
This ratio is very sensitive to the relative timings, and changes of 10~\us can change this ratio by
a factor of $\sim$2. This is due to the gas pulse directly after the valve being significantly
shorter (opening time of the piezo is around $25$~\us) than in the detection
region located $\sim$0.5~m downstream where the measurements were
taken. This is also reflected by the steep falloff of signal for later desorption laser timings.

\begin{figure}[t]
   \centering
   \includegraphics[width=\figwidth]{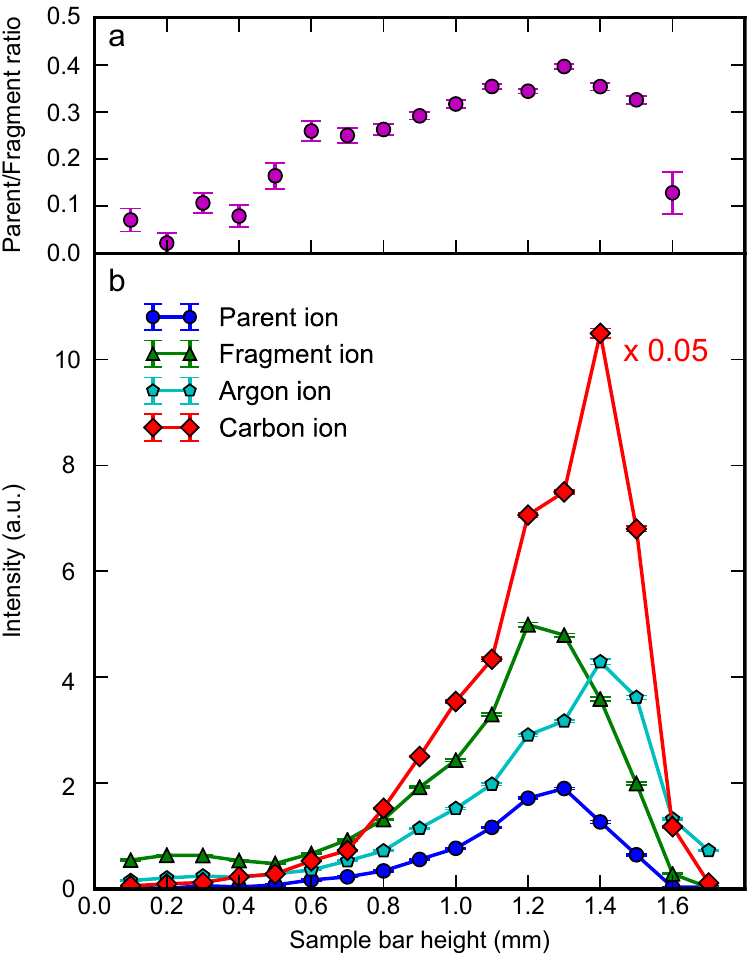}
   \caption{Measured ion intensity for parent APCN, the characteristic fragment at $m/z=162$, carbon
      and argon as a function of sample bar height. The top panel shows the ratio of parent to
      fragment ions observed.}
   \label{fig:height_scan}
\end{figure}

As shown above, the sample bar has a large effect on the supersonic expansion and hence the produced
molecular beam. To investigate this further we have taken data for different heights of the sample
bar, shown in \autoref{fig:height_scan}.
Here, we recorded the intensity of argon, carbon, fragment and parent ions for different heights of
the sample bar, as well as the ratio of detected parent to fragment ions. We note that all this data
was taken at identical timing of $t_\text{valve}=-900$~\us and $t_\text{desorption}=80$~\us and
that the height of the sample bar does change the desorption laser focusing conditions (since the
focusing lens is fixed relative to the valve).

It is evident from this data that the sample bar height changes not only the peak intensity, but
also the parent to fragment ratio, as shown in \autoref{fig:height_scan}(a). The observed
intensities are very sensitive to the height of the sample bar, with parent, fragment and carbon
ions showing maxima at different positions. This sensitivity was used in all previous measurements
to fix the height for each new sample bar measured; it was optimized prior to taking data to obtain
maximum signal from the parent ion. However, we note that due to the preparation method, slight
differences in height can persists even across a single sample bar, and affect the measurement and
especially the comparability between data sets. This could be the reason for the large deviation of
the data shown for 75~\us in \autoref{fig:peak_ratio}. This sensitivity can be explained by a number
of contributing factors; (i) the efficiency with which desorbed molecules are picked up and carried
by the argon beam. Assuming that the slight differences in mass and size between parent and
fragments are negligible given the very large number of collisions with the carrier gas, this should
be comparable for all species within the beam, \ie, a lower pressure of argon should affect all
species to a comparable extent. (ii) The degree to which the argon expansion is disturbed by the
presence of the sample bar. In \autoref{fig:beam_timing}~a we have shown that the sample bar
influences the molecular beam speed and distribution; at the fixed $t_\text{valve}$ conditions used
here this will change the observed intensities. It has furthermore been shown already that the
sample bar changes the directionality of the molecular beam~\cite{Arrowsmith:APPB2:165}. (iii)
Changes in the cooling efficiency of desorbed species lead to differences in the parent to fragment
ratio in the beam. (iv) The spotsize of the desorption laser on the sample bar changes with
different sample bar heights and thus the intensity of the laser and the number of molecules
interacting with the laser is influenced. From the collected data we cannot comment on the relative
importance of these different mechanisms, however, since (i) and (ii) should influence the parent
and fragment molecules nearly identically, the observed changes in the parent to fragment ratio
indicate a dependence of the cooling efficiency on the sample-bar height. The cooling efficiency
should be best within the densest part of the molecular beam, which is on the axis of the 300~\um
nozzle orifice. Therefore, a sample bar height just below this position, \ie, covering slightly less
than half the nozzle opening, should lead to the densest plume of desorbed molecules being entrained
in the densest part of the molecular beam. This simple consideration is consistent with our
observations of the maximum parent signal and parent to fragment ratio occurring at the position
labeled 1.3~mm. While we cannot disentangle all the different effects of changing the sample bar
height, it is clear that this, and the associated influence on the supersonic expansion and
molecular beam properties, is a crucial parameter for laser desorption entrainment of molecules into
supersonic expansions. This could be further investigated by either measuring the gas flow from the
nozzle directly, for example through direct visualization of gas
densities~\cite{Luria:JPCA115:7362,Horke:JAP121:123106} or by measuring spatial argon profiles
through strong-field-ionization mass spectrometry at various distances from the nozzle.

Despite the wide use of laser desorption sources, very few studies have looked into the fundamental
underlying processes. Furthermore, the vastly different source designs in use, \eg, different
desorption laser wavelengths, intensities, pulse durations, different models of supersonic valves,
etc., make comparison to previous studies difficult. While we believe this is the first study of
laser desorption using strong-field ionization, previous experiments have utilized electron impact
ionization and have similarly observed a large amount of neutral fragments produced by the
desorption processes~\cite{Vastola:AMS4:107}. Our finding that the molecular packet of desorbed
molecules is much shorter than the envelope of the seeding-gas pulse is also consistent with
previous measurements~\cite{Arrowsmith:APPB2:165}. Lastly, we point out that the use of a
fiber-coupled desorption laser has been demonstrated before~\cite{Piuzzi:CPL320:282}, albeit without
refocusing inside the vacuum chamber that we have introduced here for greater control.

\section{Conclusion}

We have presented a novel laser-desorption setup designed for use in advanced imaging experiments of
ultrafast molecular dynamics and we have carefully characterized and optimized the laser-desorption
and molecular-beam-entrainment conditions. The setup consists of a single central mechanical unit
containing all necessary parts (molecular beam valve, sample bar with motors and desorption laser
optics) that is mounted on an \emph{XYZ} manipulator on a single flange for independent motion.
Furthermore, we have presented a detailed characterization of our new laser-desorption source as
well as molecular beams produced using laser desorption in general. By utilizing strong-field
ionization, we were able to probe all species contained within the beam. Under normal operating
conditions we found that the molecular beam contains, in addition to parent molecules, significant
amounts of molecular fragments, as well as carbon clusters from the desorption process. We
investigated the role of the desorption laser fluence, the relative timing of valve opening and
desorption laser, the sample bar height, and which part of the molecular packet is probed. While
increased desorption laser fluence leads to more molecules contained within the molecular beam, it
was found to induce fragmentation of the sample and leads to enhanced contamination of the beam with
carbon and its clusters. The placement of the desorbed plume of molecules within the gas pulse from
the supersonic expansion has a profound effect on the cooling efficiency, and thus the fragmentation
observed. The best timing was found to be approximately in the center of the gas pulse, and is quite
sensitive compared to the gas pulse duration in the detection region. The relative height of the
sample bar in front of the valve orifice significantly affects the molecular beam expansion
conditions, and hence the intensity of observed signals, as well as the parent to fragment ratio.
However, finding the optimum position for the sample bar height is difficult, due to the number of
competing effects taking place, and every sample bar being unique. Furthermore,
   parameters might be dependent on the employed molecular-beam nozzle, and our exact findings are
   specific to the used conical nozzle shape.

From our detailed investigation we found that the optimal settings for building a laser desorption
source very much depend on the planned experimental scheme. While some parameters, such as the
relative timing of desorption laser and the molecular beam valve, should always be optimized as
shown here, other parameters are not critical. For example the pulse energy of the desorption laser
should be chosen according to the application. For techniques that are only sensitive to intact
parent molecule signal, like resonance-enhanced multiphoton ionization, the pulse energy of the
desorption laser should be kept high because this increases the number density of parent molecules
in the interaction region. But for non-species specific techniques, such as x-ray diffraction, the
pulse energy should be reduced to minimize the contamination with fragments and carbon. However,
even at the lowest desorption energy used here, we still observe a significant amount of molecular
fragments and carbon clusters in the beam. While the former originate to some extent from the
strong-field-ionization probing, the carbon and carbon clusters are certainly in the beam due to the
desorption process. In order to produce a pure beam of intact parent molecules in the gas-phase, one
can consider coupling a laser desorption source with other species separation techniques for neutral
molecules, such as electrostatic deflection or alternating gradient focusing~\cite{Chang:IRPC34:557,
   Filsinger:PRL100:133003} and such experiments are currently underway in our laboratory.

\section*{supplementary material}\noindent%
See supplementary material for detailed 3D drawings of the laser desorption source and individual
components.

\begin{acknowledgments}\noindent%
   We thank Anouk Rijs for helpful discussions regarding the design of our new source, Horst Zink
   and the DESY FS electronics workshop for expert support with the setup, and Ortwin Hellmig and
   Andreas Bick for help with the fiber vacuum feedthrough.

   In addition to DESY, this work has been supported by the European Research Council under the
   European Union's Seventh Framework Programme (FP7/2007-2013) through the Consolidator Grant
   COMOTION (ERC-614507-Küpper), by the excellence cluster ``The Hamburg Center for Ultrafast
   Imaging -- Structure, Dynamics and Control of Matter at the Atomic Scale'' of the Deutsche
   Forschungsgemeinschaft (CUI, DFG-EXC1074), and by the Helmholtz Gemeinschaft through the
   ``Impuls- und Vernetzungsfond''. We gratefully acknowledge a Kekulé Mobility Fellowship by the
   Fonds der Chemischen Industrie (FCI) for Nicole Teschmit.
\end{acknowledgments}

\bibliography{string,cmi}
\end{document}